\makeatletter
\@ifundefined{@parse@version@dash}{%
\def\@parse@version#1{\@parse@version@0#1}
\def\@parse@version@#1/#2/#3#4#5\@nil{%
\@parse@version@dash#1-#2-#3#4\@nil}
\def\@parse@version@dash#1-#2-#3#4#5\@nil{%
  \if\relax#2\relax\else#1\fi#2#3#4 }
}{}
\makeatother

\documentclass[reprint,amsmath,amssymb,aps]{revtex4-1}
\usepackage{graphicx}
\usepackage{dcolumn}
\usepackage{bm}
\usepackage{harpoon}

\begin{document}

\preprint{APS/123-QED}

\title{Searching for Exotic Spin-Dependent Interactions Using Rotationally Modulated Source Masses and an Atomic Magnetometer Array   }

\author{K.Y. Wu}
\affiliation{Key Laboratory of Neutron Physics, Institute of Nuclear Physics and Chemistry, CAEP, Mianyang, Sichuan, 621900,China }

\author{S.Y. Chen}
\affiliation{Key Laboratory of Neutron Physics, Institute of Nuclear Physics and Chemistry, CAEP, Mianyang, Sichuan, 621900,China }

\author{J.Gong}
\affiliation{Key Laboratory of Neutron Physics, Institute of Nuclear Physics and Chemistry, CAEP, Mianyang, Sichuan, 621900,China }

\author{M. Peng}
\affiliation{Key Laboratory of Neutron Physics, Institute of Nuclear Physics and Chemistry, CAEP, Mianyang, Sichuan, 621900,China }

\author{H.Yan}
\email[Corresponding author: ]{hyan@caep.cn}\affiliation{Key Laboratory of Neutron Physics, Institute of Nuclear Physics and Chemistry, CAEP, Mianyang, Sichuan, 621900,China }

\date{\today}

\begin{abstract}
We describe a proposed experimental search for exotic spin-dependent interactions using rotationally modulated source masses and an atomic magnetometer array.
Rather than further improving the magnetometer sensitivity, noise reduction can be another way to reach higher measurement precision. In this work, we propose to use modulating techniques of the source masses to reduce the noise of the experiment. Better precision can be achieved if the fundamental frequency and harmonics of the rotating source masses are used to detect the new interactions.
Furthermore, if an array of magnetometers are applied, the statistical precision can be improved, and some common-mode noises can be canceled. Our analysis and simulations indicate that the proposed experiment scheme can improve the detection precisions of three types of spin-dependent interactions by as much as $\sim$5 orders of magnitude in the force range of $\sim$cm to $\sim$10m. 
\end{abstract}

\maketitle

\section{\label{sec:level1}Introduction}
Spin-dependent new interactions beyond the Standard Model are related to the solutions to several important questions of modern physics. Exotic interactions mediated by axions are one of the examples\cite{PEC1977,WEI78,WIL78}. On one hand, axions are possible candidates for dark matter, which remains one of the most important unsolved problems in particle physics and astrophysics. On the other hand, axions are attractive in particle physics since they probably provide the most promising solution to preserve the CP-symmetry in strong interactions. The axion was initially introduced to solve the strong CP problem in QCD in which new bosons occur as a consequence of the spontaneous breaking of Pecci-Quinn symmetry\cite{PEC1977,PIE2021}.

The ALPs(Axion Like Particles), if exist, can generate a new interaction of the form  $\mathcal{L}_{\phi}=\bar{\psi}(g_{S}+ig_{P}\gamma_{5})\psi\phi$ through  a light scalar boson $\phi$ coulping to a fermion $\psi$ ,where $g_S$ and $g_P$ are the scalar and pseudo-scalar coupling constants\cite{Moody1984}. The SP(scalar-pseudoscalar) interaction or the monopole dipole interaction
has begun to attract more scientific attention recently\cite{ARV2014,TER2015,CRE2017,GER17,LEE2018,AGG20,CRE2022}. The interaction between the probe particle
of the polarized electron and the source particle of unpolarized fermion can be expressed as:
\begin{equation}\label{eqnSP}
V_{SP}(r)=\frac{\hbar^{2}g_{S}g_{P}}{8\pi m_e}(\frac{1}{\lambda r}+\frac{1}{r^{2}})\exp{(-r/\lambda)}\vec{\sigma}\cdot\hat{r},
\end{equation}
where $\lambda=\hbar/m_{\phi}c$ is the interaction range, $m_{\phi}$ is the mass of the new scalar boson, $\vec{s}=\hbar\vec{\sigma}/2$ is the spin of the
polarized electron, $m_e$ is the electron mass and $r$ is the distance between the two interacting particles.

ALPs are scalar force carriers. New forces may be mediated by vector particles such as the para-photon (dark, hidden, heavy or secluded photon)\cite{HOL1986,DOB2005}, Z' boson\cite{PDG2020}, graviphoton\cite{ATW2000}, etc., or even unparticles\cite{LIA07}. Long ago, Fayet\cite{FAY1980a,FAY1980b} pointed out that the spontaneous breaking of the supersymmetric theories would lead to a new spin-1 boson which has a small mass and very weak couplings to quarks and leptons.
 Starting from rotational invariance, Dobrescu and Mocioiu\cite{DOB06} formed 16 different operator structures involving the spin and momenta of the interacting particles.
New interactions mediated by ALPs are a subset of the new theory. Most of the new interactions are spin-dependent. The addition of the spin degree of freedom opens up a large variety of possible new interactions to search for which might have escaped detection to date. Various experiments have been performed or proposed recently to search for a subset of these new interactions which could couple to the spin of the neutron/electron.
Studies on muons have been carried out recently\cite{YAN19}.

For the vector force carriers, the interaction can be deduced from the coupling $\mathcal{L}_{X}=\bar{\psi}(g_{V}\gamma^{\mu}+g_{A}\gamma^{\mu}\gamma_{5})\psi X_{\mu}$ where $X_\mu$ is the new vector particle, $g_V$ and $g_A$ are the vector and axial coupling constants\cite{Yan2013,Yan2015,MAL16,FAD18,FAD19}. There is  the VA(vector-axial-vector) interaction $V_{VA}(r)$ ($V_{12,13}$ in Ref.\cite{DOB06}'s notation) :
  \begin{equation}\label{eqnVA}
  V_{VA}(r)=\frac{\hbar g_{V}g_{A}}{4\pi}\frac{\exp{(-r/\lambda)}}{r}\vec{\sigma}\cdot\vec{v},
  \end{equation}
where $\vec{v}$ is the relative velocity between the probe particle and source particle, $\lambda=\hbar/m_{X}c$ is the interaction range, $m_{X}$ is the mass of the new vector boson.
  $V_{VA}(r)$ is the Yukawa potential
  times the $\vec{\sigma}\cdot\vec{v}$ factor, which makes this interaction quite interesting. Another interaction requiring only one particle to be spin-polarized is
  the AA (axial-axial) interaction $V_{AA}(r)$ ($V_{4,5}$ in Ref.\cite{DOB06}'s notation), which is also originated from the $\mathcal{L}_{X}$ coupling, can be written as:
\begin{equation}\label{eqnAA}
V_{AA}(r)=\frac{\hbar^{2}g_{A}^{2}}{16\pi m_ec}(\frac{1}{\lambda r}+\frac{1}{r^{2}}) {\exp{(-r/\lambda)}}\vec{\sigma}\cdot(\vec{v}\times\hat{r}).
\end{equation}

All these interactions $V_{SP}$, $V_{VA}$ and $V_{AA}$ are in the form of $\vec{s}\cdot\vec{B'}$ where $\vec{B'}$ can be viewed as a pseudo magnetic field\cite{PIE11}. Searching for these new interactions becomes a problem of detecting the pseudo magnetic field.
The high magnetic field sensitivity based on polarized valence electrons of alkali metals makes SERF (Spin Exchange Relaxation Free) Atomic Magnetometer(AM)\cite{ALL2002,BUD2013} a convenient choice to search\cite{VAS2009,Chu2016PRD,LEE2018,Ji2018} for the exotic spin-dependent interactions for polarized electrons.
\section{\label{sec:level1}Proposed Experimental setup}
\begin{figure*}
\includegraphics[scale=1, angle=0]{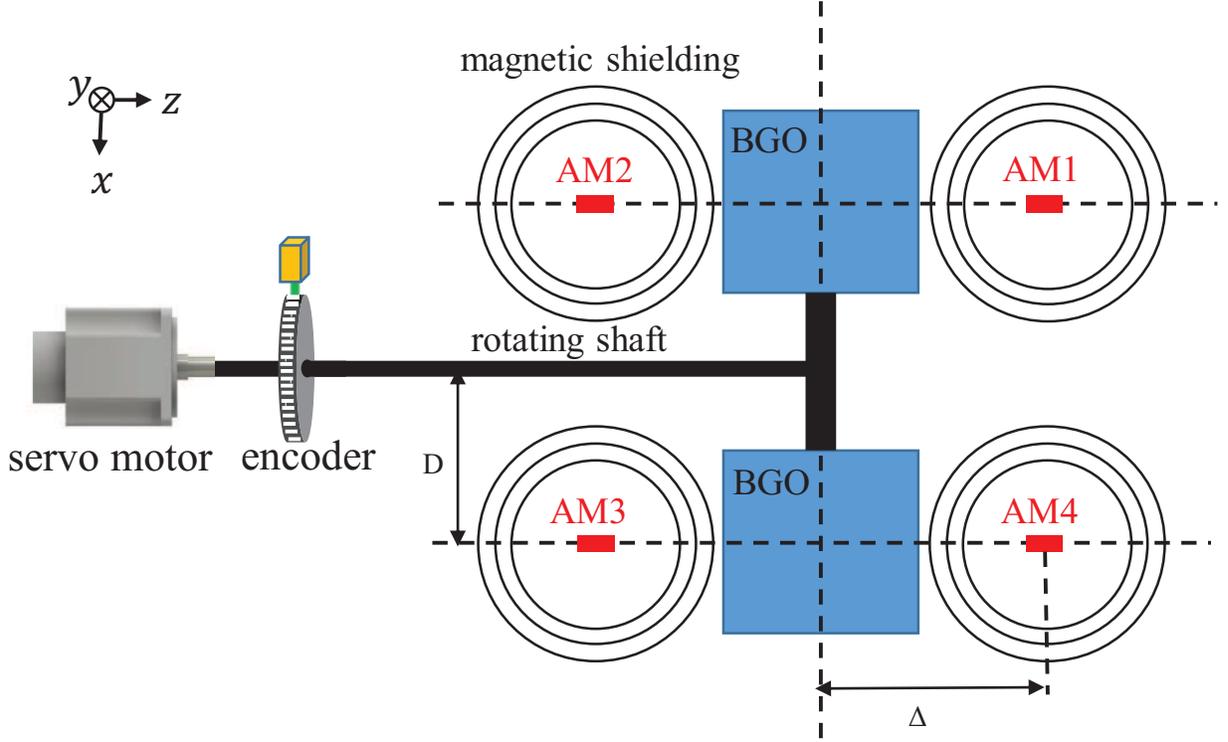}
\caption{\label{fig:setup}Schematic of the proposed experiment. The servo motor rotates two dense, nonmagnetic cylinder source masses with frequency $\omega_0$, inducing pseudo magnetic field signals to the surrounding AMs if exotic spin-dependent interactions exist. The encoder monitors the rotating angle and frequency in real-time.}
\end{figure*}

The schematic of the experimental setup is shown in Fig.~\ref{fig:setup}.
A servo motor is rotating the two identical cylinder source masses such as BGO (Bi$_4$Ge$_3$O$_{12}$) crystals.
The BGO crystal has a high number density of nucleons ($4.26 \times 10^{30} m^{-3}$) and very low susceptibility ($-19.0 \times 10^{-6}$)\cite{2003Scintillator}. BGO crystals are usually used as $\gamma$ ray scintillators. Due to its high density, high purity, and low magnetic susceptibility, it has been used as source masses in several experiments\cite{TUL2013,Kim2018,Kim2019,Ji2018} searching for the new spin-dependent interactions. The BGO crystals with a diameter of 10.16cm are commercially available and chosen as the source masses for the proposed experiment.
The rotating angle and frequency are monitored in real-time by the encoder. An array of four identical, high sensitivity, commercially available AMs with the intrinsic noise level of ~10fT.{Hz}$^{-\frac{1}{2}}$ is used to detect the new interactions. The AMs using enriched $^{87}$Rb vapor as the working medium are supposed to have a bandwidth of 200Hz, but operations for frequencies $\sim$kHz were found to be possible\cite{SAV2017}. The exotic spin-dependent interactions due to the source masses, if they exist, can induce pseudo magnetic field signals for the polarized electron spin of the surrounding AMs. Kim\cite{Kim2018,Kim2019} et al. are the pioneers of using commercially available AMs\cite{quspin} to study $V_{VA}$ and $V_{AA}$ for spin-polarized electrons.
The commercial AM\cite{quspin} has a relatively lower sensitivity, but they are compact and can be easily arranged in an array of 50 to detect the magnetic field of the human brain\cite{BOT2018,BOT2021,REA2021}. The array form has never been used to search for the spin-dependent new interactions to the best of our knowledge.

\begin{figure}[b]
\includegraphics[scale=0.65]{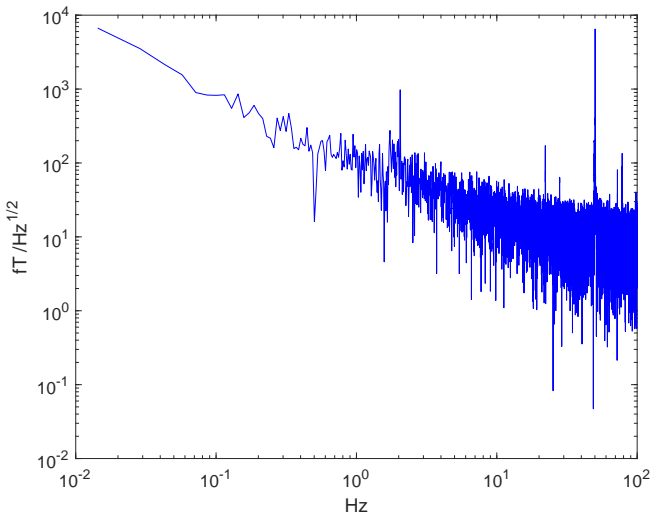}
\caption{\label{fig:PSD}Typical noise spectral density of the AM. Notice the 1/f noise feature.}
\end{figure}

The rotating frequency is assumed to be 10Hz, and modulating frequency for the source masses is 20Hz since they are in a symmetric arrangement. The benefit of modulating the masses to a frequency as high as 20Hz is apparent when looking into the noise spectral density of the Atomic Magnetometer(FIG.\ref{fig:PSD}). The noise level can be reduced by orders of magnitude. Thus, the SNR (Signal to Noise Ratio) increases accordingly with the higher frequency.

The advantage of using an array of AMs can be explained as follows. Suppose the SP type new interaction V$_{SP}$ is under search and B$'_{SPz}$ the induced pseudo magnetic field along with $\hat{z}$ direction. The AMs configured as an array in Fig.~\ref{fig:setup} have different responses for the induced pseudo magnetic field. AM1 and AM4 will sense a signal along $+\hat{z}$ direction while AM2 and AM3 $-\hat{z}$ direction.
The signal due to the new interaction can be extracted as:
\begin{eqnarray}
B'_{SPz}=\frac{1}{4}(AM1_z-AM2_z-AM3_z+AM4_z).
\end{eqnarray}
If there is some kind of common noise, it can be largely canceled. On the other hand, statistics can increase due to using an array of AMs.
\section{\label{sec:level1}Data processing method}

\begin{figure}[htbp]
 \includegraphics[scale=0.325, angle=0]{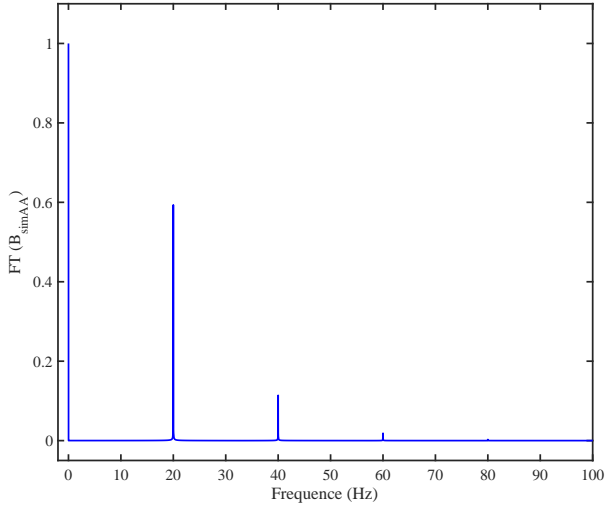}
 \caption{\small{Typical Fourier spectrum of the simulated signal generated by the AA type interaction along the most sensitive direction. $c_n$'s are normalized to $c_0$. }}
 \label{fig.FTAA}
\end{figure}

As the source masses rotate at a constant speed, periodic signals due to new interactions are supposed to be generated. It is natural to choose a data processing method based on Fourier analysis. When taking $g_Sg_P=g_Vg_A=g_Ag_A=1$, theoretically, the pseudo magnetic field at the point $\vec{r}$ can be calculated and expressed as:
\begin{widetext}
\begin{eqnarray*}
\label{B'}
\vec{B'}_{SP}(\vec{r}) =\frac{ \hbar g_{S}g_{P}}{4\pi m_e\gamma_e}\int d^3\vec{r}'(\frac{1}{\lambda |\vec{r}-\vec{r}'|}+\frac{1}{|\vec{r}-\vec{r}'|^{2}})
{\exp{(-|\vec{r}-\vec{r}'|/\lambda)}}(\frac{\vec{r}-\vec{r}'}{|\vec{r}-\vec{r}'|}),\\
\vec{B'}_{VA}(\vec{r}) =\frac{ g_{V}g_{A}}{2\pi\gamma_e}\int d^3\vec{r}'\frac{\exp{(-|\vec{r}-\vec{r}'|/\lambda)}}{|\vec{r}-\vec{r}'|}\vec{v},\\
 \vec{B'}_{AA}(\vec{r})=\frac{\hbar g_{A}^{2}}{8\pi m_ec\gamma_e}\int d^3\vec{r}'(\frac{1}{\lambda |\vec{r}-\vec{r}'|}+\frac{1}{|\vec{r}-\vec{r}'|^{2}})
 {\exp{(-|\vec{r}-\vec{r}'|/\lambda)}}(\vec{v}\times\frac{\vec{r}-\vec{r}'}{|\vec{r}-\vec{r}'|}),
\end{eqnarray*}
\end{widetext}
where  $\gamma_e$ is the gyromagnetic ratio of the electron, $d^3\vec{r}'$ is a three-dimensional volume element at $\vec{r}'$ of the source mass. The probing polarized particle is assumed to be the electron since the AM uses polarized electrons. The above integrations can be calculated using techniques such as the Monte Carlo integration method\cite{PRE2007}. A total of $10^6$ random points are sampled both in the source mass and the AM cell to perform the Monte Carlo integration. Although sampling of $10^7$ points is also compatible with our computing power, the error for $10^6$ points is found to be within 1\%, which is similar with Ref.\cite{Kim2018,Kim2019} and good enough for our purpose.  
Then the result can be expanded as Fourier series:
\begin{eqnarray}
\label{B'}
\nonumber B'(t) &=& {c_0} + {c_1}\cos ({\omega _0}t) + {c_2}\cos (2{\omega _0}t) + {c_3}\cos (3{\omega _0}t) \\
  & & + {c_4}\cos (4{\omega _0}t) + ...,
\end{eqnarray}
where $\omega _0 = 2\pi {f_0}$  and ${f_0=20Hz}$ is modulating frequency of the source masses. For simplicity and without losing generality, here we only considered the cosine terms of the Fourier series. It is reasonable to make this simplification since the initial angular position or the phase of the system, in principle, can be set before taking measurements to make the expansion only has cosine terms. $c_n$'s can be expressed as:
\begin{eqnarray}
{c_n} = \frac{{\int_0^{T}  {\cos \left( {n{\omega _0}t} \right){B'(t)}dt} }}{{\int_0^T  {{{\cos }^2}\left( {n{\omega _0}t} \right)dt} }}.
\end{eqnarray}
The typical Fourier spectrum of the simulated signal for the AA type interaction is shown as Fig.\ref{fig.FTAA}. Similar results are observed for the simulated signal of SP and VA interactions.

In actual experiments, the observed signal is supposed to be:
\begin{eqnarray}
\label{Bexp1}
{B_{\exp }}(t) = \alpha{B'(t)}+n(t),
\end{eqnarray}
where $\alpha$ is the actual strength of the new interactions, i.e. $\alpha=g_Sg_P$ for the SP type interaction, $\alpha=g_Vg_A$ for the VA type interaction and $\alpha=g_Ag_A$ for the AA type interaction respectively. $n(t)$ is the noise. Again, expand $B_{\exp}(t)$ in Fourier series with fundamental frequency $\omega_0$, we will have:
\begin{eqnarray*}
\label{B'}
\nonumber B_{\exp}(t) = \alpha{c_0} + \alpha{c_1}\cos ({\omega _0}t) + \alpha{c_2}\cos (2{\omega _0}t)\\+ \alpha{c_3}\cos (3{\omega _0}t)
 + \alpha{c_4}\cos (4{\omega _0}t) + ...+n(t).
 \end{eqnarray*}
Now, $\alpha$ the interaction coupling constant can be extracted from the measurements as:
\begin{equation}
\alpha|_n= \frac{{\int_0^T  {\cos \left( {n{\omega _0}t} \right){B_{\exp}(t)}dt} }}{{c_n\int_0^T  {{{\cos }^2}\left( {n{\omega _0}t} \right)dt} }},
\end{equation}
where the noise will contribute as:
\begin{equation}
\delta\alpha|_n= \frac{{\int_0^T  {\cos \left( {n{\omega _0}t} \right){n(t)}dt} }}{{c_n\int_0^T  {{{\cos }^2}\left( {n{\omega _0}t} \right)dt} }}.
\end{equation}
The upper integration limits of Eqn.(6),(8) and (9) are taken to be $T$ in this paper. In practice, an integer number of periods is supposed to be used. It is easy to show that the method works the same way in this practical case. Assume the actual integration time, T is large enough, and the noise contribution can be estimated as\cite{LIB2003,YAN2014}:
\begin{equation}
\delta\alpha|_n\sim \frac{\sqrt{2}}{c_n}\sqrt{S_N(nf_0)}\sqrt{\frac{1}{T}},
\end{equation}
where $S_N(nf_0)$ is the noise power density at frequency $n\omega_0$. The integration acting as a low pass filter reduces the noise bandwidth, thus increasing the SNR of the measurement.

In principle, all the terms in the Fourier expansion can be used to determine $\alpha$. Terms with large $c_n$s will be disturbed less by the same noise level. Thus the weighted average method should reduce the noise and obtain better statistics. Furthermore, as it can be seen from FIG.\ref{fig:PSD} and FIG.\ref{fig.FTAA}, to avoid the $1/f$ noise, the DC or the $c_0$ term should not be used . Taking into account the actual bandwidth of the AM, the interaction strength can finally be determined as,
\begin{eqnarray}
\bar{\alpha}= \frac{{\sum\limits_{n = 1}^4 {c_n^2{\alpha|_n}} }}{{\sum\limits_{n = 1}^4 {c_n^2} }}.
\end{eqnarray}
As seen in FIG.2, the noise power densities at the interested frequencies are at the same level. It is easy to derive that,
\begin{equation}
\delta\bar{\alpha}|\sim \sqrt{2S_N(nf_0)}\sqrt{\frac{1}{T}}\frac{1}{\sqrt{c_1^2+c_2^2+c_3^2+c_4^2}},
\end{equation}
which is better than using the single frequency of either fundamental or harmonics.
\begin{table}[htbp]
\caption{\label{tab:table3}%
Parameters used in the simulation. }
\begin{ruledtabular}
\begin{tabular}{llrrr}
Parameter&Value&
\\
\hline
source mass density(BGO) & 7.13g.cm$^{-3}$ \\
number density of nucleons (BGO) & 4.26$\times$10$^{24}$cm$^{-3}$ \\
source mass diameter (cylinder) &  10.16cm \\
source mass height & 10.16cm \\
noise level of the AM  & 15fT.Hz$^{-\frac{1}{2}}$ \\
alkali vapor cell size of the AM & 0.3$\times$0.3$\times$0.3cm$^3$ \\
rotating frequency, f & 10Hz \\
rotating radius, r & 10cm \\
distance between the source and sensor, $\Delta$ & 10cm\\
single simulation run time duration,  & 60s \\
total number of simulation runs,  & 43200\\
\end{tabular}
\end{ruledtabular}
\end{table}

\section{\label{sec:level1}Projected sensitivity}

\begin{figure}[htbp]
  \centering
   \includegraphics[scale=0.45, angle=0]{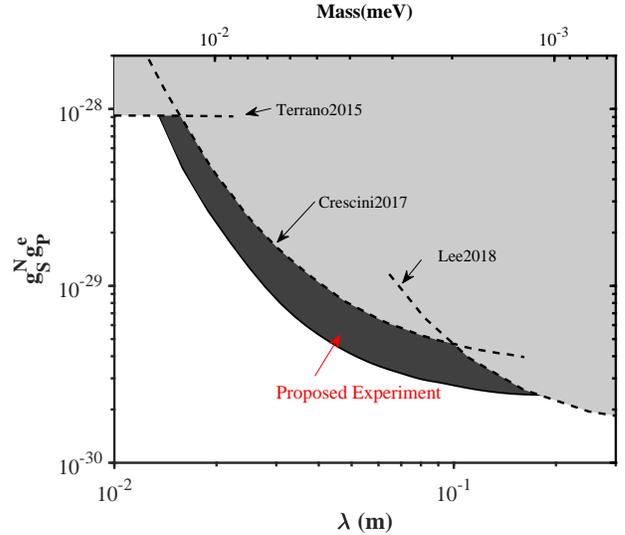}
   \caption{\small{Expected 2$\sigma$ sensitivity (solid line) of the proposed experiment for the SP interaction. The light grey area is the excluded area
   by present experiments. The dashed line is the result of \cite{CRE2017,TER2015,LEE2018}. Here we used the $2\sigma$ limit to be consistent with the relevant references. }}
   \label{fig.gsgp}
  \end{figure}
 
Several features of the proposed experiment can reduce the noise, thus increasing the SNR of the measurements. Improvement on sensitives is expected, and Monte Carlo simulations are applied to check if it is the case. The parameters used in the simulations are listed as TABLE I. Every run of the measurements is performed in a time window of 60s. The total run number is 43200, thus resulting in a total integration time of 30 days.

\begin{figure}[htbp]
  \centering
   \includegraphics[scale=0.45, angle=0]{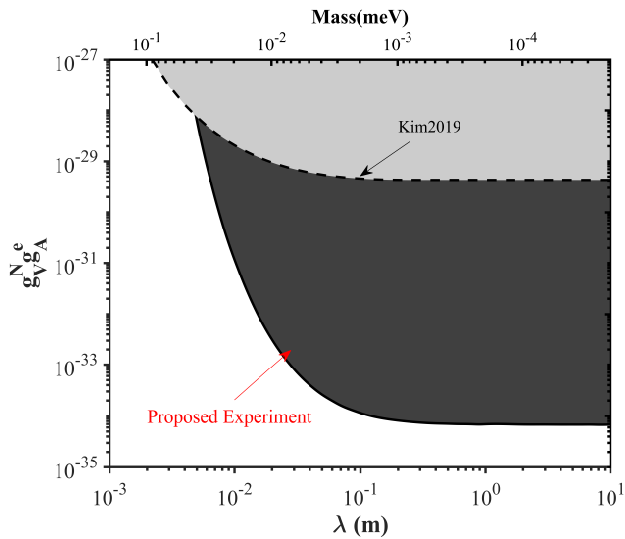}
   \caption{\small{Expected 1$\sigma$ sensitivity(solid line) of the proposed experiment for the VA interaction. The light grey area is the excluded area
   by present experiments. The dashed line is the result of \cite{Kim2018}. Here we used the $1\sigma$ limit to be consistent with the relevant reference.  }}
   \label{fig.gvga}
  \end{figure}

\begin{figure}[htbp]
  \centering
   \includegraphics[scale=0.45, angle=0]{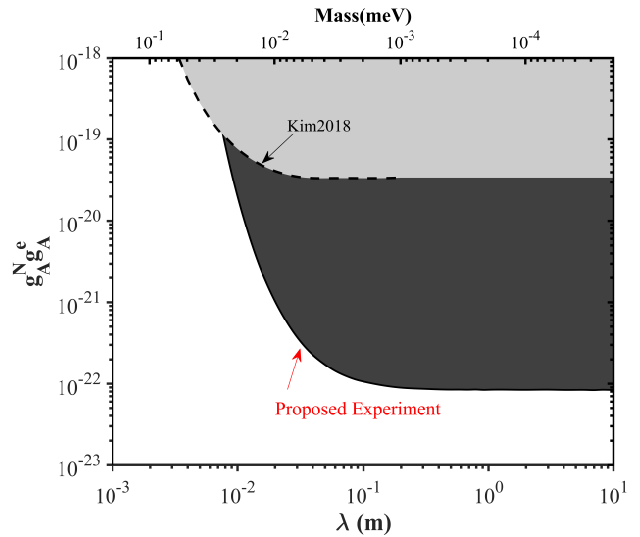}
   \caption{\small{Expected 1$\sigma$ sensitivity(solid line) of the proposed experiment for the AA interaction. The light grey area is the excluded area
   by present experiments. The dashed line is the result of \cite{Kim2019}.Here we used the $1\sigma$ limit to be consistent with the relevant reference.}}
   \label{fig.gaga}
  \end{figure}

The data processing procedure is as follows.  With the known $\omega_0$, we firstly calculate $B'(t)$ for the specfic $\lambda$ using Monte Carlo techniques, then $c_n$ can be obtained by numerically integrating Eqn.(6), as described in Section III. Using $c_n$ obtained  previously, $\alpha$ or $g_Sg_P$, $g_Vg_A$ and $g_Ag_A$, can be calculated by integration of Eqn.(8) using $B_{\exp}(t)$ time-series generated by the Monte Carlo simulations. Simpson's method, which is a numerical integration technique with high precision\cite{YAN2014,PRE2007}, is applied throughout the work. Repeat the procedure for different $\lambda$ points, and we obtain $g_Sg_P$, $g_Vg_A$, and $g_Ag_A$ for the interested interaction range.
The expected sensitivities for the SP, VA and AA interactions are shown in FIG.\ref{fig.gsgp},\ref{fig.gvga} and \ref{fig.gaga}.  As much as $\sim$5 orders of magnitude improvement is obtained for $g_V^Ng_A^e$(where "N" stands for the nucleon and "e" for electron) and  $\sim$3 orders of magnitude for $g_A^Ng_A^e$ in the force ranges of $\sim$0.01m to 10m. The sensitivity for $g_S^Ng_P^e$ is also expected to be improved in the range of $\sim$0.01m to 1m.
\section{\label{sec:level1}Conclusion and Discussion}
This paper proposes a new experimental scheme to detect the exotic spin-dependent interactions of SP, VA, and AA types. Rather than doing the mass in and out operation, we propose to modulate the source masses to a frequency as high as 20Hz. A data processing strategy based on the Fourier series is described. The DC term is omitted to avoid the $1/f$ noise. The fundamental frequency term and several harmonics are used in the weighted average way to determine the modulated signal. Technically, the data processing is based on the integration method; thus, high-frequency noise can be reduced\cite{YAN2014}. Monte Carlo simulations are applied to verify the validity of the proposed experiment. Sensitivities on SP, VA, and AA type interactions are expected to be improved by as much as $\sim$5 orders of magnitude in the range of $\sim$0.01m to $\sim$10m. 
For carrying out the experiments, systematics due to vibrations caused by rotating the two $\sim$6Kg source masses at 600RPM are the biggest concerns. Our initial tests indicate that the strong signal due to the mechanical coupling of vibrations shows up. Thus,  we must apply vibration isolation techniques to perform reasonable measurements. On the other hand, Ref.\cite{SU2021} reported significant systematic effects caused by air currents or air vibrations which are also due to rotations of the source masses. It seems we should prepare to use the necessary shieldings to avoid that effects too. Other factors such as the rotating frequency precision, initial phase uncertainty, Monte Carlo integration error, etc., were also considered. We found that the uncertainty due to these effects is at least one order of magnitude less than the aimed precisions. 
According to the proposed scheme, the experiment has already started, and the results are expected to be obtained soon.

We acknowledge support from the National Key Program for Research and Development of China, under grant 2020YFA0406001 and 2020YFA0406002. This work was also supported by the National Natural Science Foundation of China (Grant U2030209). We thank Chongqing Medical University for the loan of the AMs.



\end{document}